\RequirePackage[table]{xcolor}
\documentclass[journal,twoside,web]{ieeecolor}
\usepackage{generic}
\usepackage{cite}
\usepackage{amsmath,amssymb,amsfonts}
\usepackage[linesnumbered,ruled,vlined]{algorithm2e}

\SetCommentSty{mycommfont}
\SetKwInput{KwInput}{Input}
\SetKwInput{KwOutput}{Output}
\usepackage{graphicx}
\usepackage{adjustbox}
\usepackage{textcomp}
\usepackage{booktabs}
\usepackage{makecell}
\usepackage{array}
\usepackage{multirow}
\usepackage{pifont}
\usepackage{tabularx}
\usepackage{float}
\usepackage{xspace}
\usepackage[colorlinks,linkcolor=blue,citecolor=blue,urlcolor=blue]{hyperref}
\hypersetup{pdftitle={X-SPUR: Explainable Surprisal-Based Protocol-Aware Unsupervised Reasoning for Automotive Ethernet Intrusion Detection}}

\usepackage[nameinlink]{cleveref}
\crefname{figure}{Fig.}{Figs.}
\crefname{algorithm}{Algorithm}{Algorithms}
\Crefname{algorithm}{Algorithm}{Algorithms}
\crefname{algocf}{Algorithm}{Algorithms}
\Crefname{algocf}{Algorithm}{Algorithms}
\crefname{table}{Table}{Tables}
\crefformat{section}{#2\S#1#3}
\crefformat{subsection}{#2\S#1#3}
\crefformat{subsubsection}{#2\S#1#3}
\crefname{equation}{}{}

\newcommand{\BfPara}[1]{{\noindent\bf#1.}\xspace}

\newcommand{\cmark}{\ding{51}} 
\newcommand{\xmark}{\ding{55}} 
\newcommand{\dmark}{\ding{108}} 

\newcolumntype{C}[1]{>{\centering\arraybackslash}p{#1}}

\definecolor{zebragrey}{gray}{0.93}

\def\BibTeX{{\rm B\kern-.05em{\sc i\kern-.025em b}\kern-.08em
    T\kern-.1667em\lower.7ex\hbox{E}\kern-.125emX}}
\begin{document}
\title{X-SPUR: Explainable Surprisal-Based Protocol-Aware Unsupervised Reasoning for Automotive Ethernet Intrusion Detection}

\author{Jisoo Kim,~\IEEEmembership{Student Member,~IEEE,} and Seonghoon Jeong,~\IEEEmembership{Member,~IEEE}
\thanks{This article has been accepted for publication in \textit{IEEE Transactions on Industrial Informatics}. This is the author's accepted manuscript.}
\thanks{Manuscript received July 13, 2026; revised September 6, 2026; accepted September 11, 2026. This research was supported by Sookmyung Women's University Research Grants (1-2403-2028). \textit{(Corresponding author: Seonghoon Jeong.)}}
\thanks{Jisoo Kim is with the Department of Data Science, Sookmyung Women's University, Seoul 04310, Republic of Korea (e-mail: sallysooo@sookmyung.ac.kr).}
\thanks{Seonghoon Jeong is with the Division of Artificial Intelligence Engineering, Sookmyung Women's University, Seoul 04310, Republic of Korea (e-mail: seonghoon@sookmyung.ac.kr).}
\thanks{\textcopyright~2026 IEEE. Personal use of this material is permitted. Permission from IEEE must be obtained for all other uses, in any current or future media, including reprinting/republishing this material for advertising or promotional purposes, creating new collective works, for resale or redistribution to servers or lists, or reuse of any copyrighted component of this work in other works.}}

\maketitle
\begin{abstract}
Automotive Ethernet carries heterogeneous multi-protocol traffic in modern in-vehicle networks, where labeled attack data are rarely available and the strongest prior unsupervised detector still relies on handcrafted traffic features. This article presents X-SPUR, an explainable, surprisal-based, protocol-aware unsupervised reasoning framework that instead represents raw packet fields as token sequences, learns benign traffic patterns through causal language modeling, and detects anomalies from per-token cross-entropy surprisal. To incorporate temporal context, we introduce a bimodal fusion architecture that combines payload-token embeddings with inter-packet timing through additive fusion and a Hadamard interaction. To handle the heterogeneous score distributions of different protocol families, we further propose a dual top-$k$\% per-protocol $Z$-score calibration that jointly captures moderately distributed and sparse anomaly signatures. On the TOW-IDS dataset, X-SPUR achieves an AUC of 0.9987. This is marginally higher than the 0.9969 reported for AERO. X-SPUR also eliminates handcrafted feature engineering. We train a separate CarDS model using the same architecture and training hyperparameters. This model retains strong performance on the second automotive Ethernet dataset. Beyond detection, per-token surprisal provides fine-grained explainability by attributing anomaly scores to specific protocol fields, supporting interpretable security analysis in heterogeneous in-vehicle networks.
\end{abstract}


\section{Introduction}
\label{sec:introduction}
\IEEEPARstart{A}{utomotive} Ethernet is rapidly replacing the legacy Controller Area Network (CAN) as the backbone of modern in-vehicle networks (IVNs), supporting heterogeneous traffic such as Audio/Video Transport Protocol (AVTP, IEEE~1722) streams, generalized Precision Time Protocol (gPTP, IEEE~802.1AS) synchronization, and service-oriented communication over UDP. As this transition increases bandwidth and protocol diversity, it also expands the attack surface of connected and autonomous vehicles. Intrusion detection systems (IDSs) for automotive Ethernet must therefore operate under heterogeneous protocol semantics while remaining effective even when labeled attack data are unavailable during deployment.

Among recent unsupervised approaches, AERO~\cite{Jeong2024AERO} has demonstrated strong detection performance by engineering three multimodal features (an abstract protocol sequence, raw payloads, and timestamp statistics) and training a lightweight neural outlier detector on benign traffic. Such feature engineering, however, introduces two limitations. First, handcrafted features may restrict generalizability across protocol variants and traffic conditions. Second, even when detection performance is strong, an outlier score computed on engineered features does not reveal which packet fields triggered an alert, limiting interpretability for security analysts. 

To address these limitations, we propose X-SPUR, an explainable unsupervised IDS for automotive Ethernet based on causal next-token prediction. X-SPUR treats parsed packet fields as a structured token sequence and learns benign traffic patterns through language modeling. Under this formulation, anomalous traffic manifests as tokens with unexpectedly high cross-entropy, or surprisal, relative to the benign distribution learned by the model. Because anomaly evidence is computed at the token level, it can be traced back to individual protocol fields, providing an intrinsic basis for explainability. 

X-SPUR is built on a Mamba2 state-space language model, whose sequence-processing cost grows linearly with input length. Raw protocol fields extracted with \texttt{tshark} are converted into text-like packet representations and tokenized using byte-level byte-pair encoding (BBPE), preserving fine-grained protocol content without requiring manual feature design. To incorporate temporal information, we introduce a bimodal fusion mechanism that combines payload-token embeddings with inter-packet timing embeddings through additive fusion and an element-wise Hadamard interaction.

A central challenge is that benign score distributions differ sharply across protocol families. This makes a global threshold on raw scores difficult to apply across heterogeneous traffic. X-SPUR therefore computes window-level anomaly scores from the top-$k$\% token surprisal values, smooths them within each flow, and normalizes them with per-protocol $Z$-score calibration. Different top-$k$ granularities favor different anomaly morphologies. The final hybrid score therefore fuses two calibrated granularities ($k=5\%$ and $3\%$).

We evaluate X-SPUR on the TOW-IDS automotive Ethernet dataset. It achieves an area under the receiver operating characteristic curve (AUC) of 0.9987. This is marginally higher than the 0.9969 reported for AERO~\cite{Jeong2024AERO}. X-SPUR operates on raw tokenized packets and also provides per-field anomaly attribution. We separately train a CarDS model using the same architecture and training hyperparameters. It achieves an AUC of 0.9923, whereas a faithful AERO reimplementation achieves an AUC of 0.6239. The results also show that protocol-aware calibration suppresses protocol-specific false positives. Token-level surprisal further yields meaningful attack signatures across different intrusion types.

The main contributions of this article are threefold:
\begin{itemize}
    \item We propose X-SPUR, an explainable unsupervised automotive Ethernet IDS that replaces handcrafted multimodal feature engineering with end-to-end causal next-token prediction over tokenized packet fields, realized by a bimodal Mamba2 architecture that fuses payload content and inter-packet timing through additive fusion and a Hadamard interaction. On the TOW-IDS dataset, X-SPUR achieves an AUC of 0.9987. AERO~\cite{Jeong2024AERO} reports 0.9969.
    \item We propose a dual top-$k$\% per-protocol $Z$-score calibration strategy that accounts for heterogeneous protocol score distributions and differing attack morphologies. We apply the same model architecture and training hyperparameters to a separately trained CarDS model. We do not conduct a CarDS-specific search over those training hyperparameters.
    \item We show that per-token cross-entropy surprisal provides intrinsic, field-level explainability, exposing distinct attack signatures and tracing both detections and residual false positives to the same protocol fields.
\end{itemize}

The remainder of this article is organized as follows: \cref{sec:background} reviews related work; \cref{sec:problem_setup} describes the detection task, dataset, and packet-to-token pipeline; \cref{sec:proposed_framework} presents the X-SPUR framework; \cref{sec:experiments} reports the experimental setup, results, and analysis; and \cref{sec:conclusion} concludes this article.

\section{Background}
\label{sec:background}

\subsection{Intrusion Detection in Automotive Ethernet}
IVN security research has traditionally targeted the CAN bus~\cite{tariq2020canadf,Jeong2024xcanids}. However, the bandwidth demands of connected and autonomous vehicles have established automotive Ethernet as the \textit{de facto} standard for next-generation IVNs, shifting the defensive problem from a single broadcast bus to switched, multi-protocol traffic that CAN-specific detectors cannot cover.

One line of automotive Ethernet IDSs relies on explicit protocol knowledge or engineered features.
Rule-based designs pair protocol-conformance checks with gated recurrent unit (GRU) analysis or decision trees~\cite{luo2023multilayerrulebasedIDS,gail2023decisiontreeIDS}.
Weighted-histogram statistics enable lightweight per-packet screening~\cite{wang2025weightedhistogramIDS}.
Feature-based detectors operate on handcrafted Scalable service-Oriented MiddlewarE over IP (SOME/IP), AVTP, or wavelet features~\cite{kim2026xgboostsomeip,carmo2022mlxgboost,Han2023TOWIDS} (\cref{tab:related_work_comparison}).

A second line employs deep learning (DL) end to end.
Convolutional neural networks (CNNs) and recurrent neural networks (RNNs) learn directly from stacked payload bytes or sequential SOME/IP traffic~\cite{jeong2021convolutional,Alkhatib2021someip,ZHANG2025residualattention}.
Supervised training presumes labeled attack traffic that zero-day threats by definition lack.
Unsupervised and semi-supervised alternatives include reconstruction-based autoencoders and sequence-to-sequence long short-term memory (LSTM) models~\cite{Alkhatib2022AVTPnetCA,Liu2024SemiSupervisedIDS,LEANDRO2025seqwatch}.
Among these, AERO~\cite{Jeong2024AERO} attains real-time unsupervised detection with a lightweight neural outlier detector over three multimodal features.

\cref{tab:related_work_comparison} summarizes these studies. Two gaps persist across them. First, detection quality still hinges on engineered inputs: even the strongest unsupervised design depends on handcrafted multimodal features, which can tie a detector to the protocol mix it was engineered for. Second, intrinsic \textit{explainability} remains confined to shallow rule-based designs: DL-based detectors localize anomalies no finer than the packet level and leave their decisions opaque to analysts. These gaps motivate the two pillars of X-SPUR---end-to-end causal modeling of raw packet tokens on a linear-time sequence backbone, and anomaly evidence attributable to individual protocol fields by construction.

\begin{table*}[t]
    \caption{Comparison of representative studies on automotive Ethernet IDS with our study. Sequence modeling: \cmark~explicit sequence model, \dmark~partial (temporal context only, {\upshape\textit{e.g.,}}~windowing or payload stacking), \xmark~none.}
    \label{tab:related_work_comparison}
    \centering
    \footnotesize
    \setlength{\tabcolsep}{4pt}
    \renewcommand{\arraystretch}{1.0}
    \begin{tabular}{C{1.6cm} C{2.6cm} C{2.8cm} C{2.8cm} C{1.3cm} C{2.5cm} C{2.2cm}}
        \toprule
        \makecell{\textbf{Study}} &
        \makecell{\textbf{Input /}\\\textbf{Modality}} &
        \makecell{\textbf{Method}} &
        \makecell{\textbf{Paradigm}} &
        \makecell{\textbf{Sequence}\\\textbf{Modeling}} &
        \makecell{\textbf{Localization}\\\textbf{Granularity}} &
        \makecell{\textbf{Explainability}} \\
        \midrule

        \cite{luo2023multilayerrulebasedIDS} (2023) &
        SOME/IP &
        Rule-based / GRU layer &
        Rule-based / Supervised &
        \cmark &
        Sequence level &
        Limited \\

        \rowcolor{zebragrey}
        \cite{wang2025weightedhistogramIDS} (2025) &
        AVTP-byte &
        Weighted histogram&
        \textbf{Unsupervised} / Statistical &
        \dmark &
        Packet level &
        Limited \\

        \cite{gail2023decisiontreeIDS} (2023) &
        SOME/IP&
        Decision-tree &
        Supervised &
        \xmark &
        Field level &
        Intrinsic \\

        \rowcolor{zebragrey}
        \cite{kim2026xgboostsomeip} (2026) &
        SOME/IP &
        XGBoost&
        Supervised &
        \dmark &
        Packet level &
        Limited \\

        \cite{carmo2022mlxgboost} (2022) &
        AVTP-byte &
        XGBoost &
        Supervised &
        \dmark &
        Window level &
        None \\

        \rowcolor{zebragrey}
        \cite{Han2023TOWIDS} (2023) &
        AVTP, gPTP, CAN/UDP &
        Wavelet + Deep CNN &
        Supervised &
        \dmark &
        Window level &
        None \\

        \cite{ZHANG2025residualattention} (2025) &
        AVTP, gPTP, UDP, SOME/IP &
        CNN, Residual attention &
        Supervised &
        \cmark &
        Window level &
        None \\

        \rowcolor{zebragrey}
        \cite{jeong2021convolutional} (2021) &
        AVTP-byte &
        2D-CNN &
        Supervised &
        \dmark &
        Packet level &
        None \\

        \cite{Alkhatib2021someip} (2021) &
        SOME/IP&
        RNN &
        Supervised &
        \cmark &
        Sequence level &
        Limited \\

        \rowcolor{zebragrey}
        \cite{Alkhatib2022AVTPnetCA} (2022) &
        AVTP-byte&
        Convolutional Autoencoder &
        \textbf{Unsupervised} &
        \dmark &
        Window level &
        None \\

        \cite{Liu2024SemiSupervisedIDS} (2024) &
        Packet fields &
        Autoencoder &
        Semi-supervised &
        \dmark &
        Window level &
        None \\

        \rowcolor{zebragrey}
        \cite{Jeong2024AERO} (2024) &
        AVTP, gPTP, CAN/UDP (multimodal) &
        Neural outlier detector &
        \textbf{Unsupervised} &
        \dmark &
        Window level &
        None \\

        \cite{LEANDRO2025seqwatch} (2025) &
        AVTP, gPTP, CAN/UDP &
        Seq2seq / LSTM &
         \textbf{Unsupervised} &
        \cmark &
        Sequence level &
        None \\

        \rowcolor{zebragrey}
        \cite{DALUZ2024multistageAE} (2024) &
        AVTP, gPTP, CAN/UDP  &
        Random Forest / Pruned CNN &
        Supervised &
        \dmark &
        Window level &
        None \\

        \textbf{Ours} &
        \textbf{Packet-field tokens} &
        Mamba-based &
        \textbf{Unsupervised} &
        \cmark &
        \textbf{Field level} &
        \textbf{Intrinsic} \\

        \bottomrule
    \end{tabular}
\end{table*}

\subsection{Sequence Modeling: From Transformers to State-Space Models}
Transformer-based architectures have achieved state-of-the-art performance in network IDSs by capturing long-range dependencies through self-attention~\cite{Wu2022RTIDS}. Dense self-attention has $O(L^{2})$ computational complexity in the sequence length $L$. This growth can increase the computational burden of long traffic contexts on automotive electronic control units (ECUs).

State-space models (SSMs), most prominently Mamba~\cite{gu2024mamba} and its successor Mamba2~\cite{dao2024mamba2}, alleviate this bottleneck by replacing attention with a selective state-space recurrence that runs in $O(L)$ time. In network security, Mamba-based detectors model long traffic sequences with Transformer-level accuracy at markedly lower computational cost~\cite{yu2025CBMAD}, and NetMamba~\cite{wang2024netmamba} showed that a unidirectional Mamba encoder suffices for efficient traffic classification.

These results, however, concern conventional IP networks; to our knowledge, SSMs remain largely unexplored for automotive Ethernet. X-SPUR closes this gap with a Mamba2-based causal language model as its detection backbone.

\subsection{Explainability in IVN Security}
To mitigate the opacity of DL-based IDSs, recent studies attach post-hoc explainable artificial intelligence (XAI) methods such as SHAP and LIME~\cite{Neupane2022XIDSsurvey}. An operator who cannot see which inputs drove an alert has little basis to trust it, debug detector mistakes, or reduce false positives~\cite{Han2021DeepAID}, and such post-hoc explanations often resolve only to whole flows or graph edges rather than individual fields~\cite{Kaya2024XCBA}. These methods probe an already-trained detector, and the added per-alert computation can bottleneck real-time in-vehicle deployment. Foundation models for traffic understanding (\textit{e.g.,}~Lens~\cite{li2026lens}) face a parallel barrier: full-scale generative inference exceeds the memory and latency budgets of automotive ECUs.

X-SPUR instead makes explanation intrinsic to detection: because its causal model scores every token by surprisal against learned benign behavior, the quantity that raises an alert also localizes it to a specific protocol field and timestep, without auxiliary explainer modules. 
Deriving explanations from the detector's own anomaly evidence is established in general network and log anomaly detection~\cite{Nguyen2019GEE,Du2017DeepLog,guo2021logbert}.
Within IVNs, X-CANIDS ranks CAN signals by per-signal reconstruction error~\cite{Jeong2024xcanids}.
X-SPUR extends this principle to protocol-field attribution in automotive Ethernet.

\section{Problem Setup and Data Representation}
\label{sec:problem_setup}

\subsection{Detection Task and Threat Setting}
\label{subsec:task_setting}

We consider unsupervised intrusion detection on automotive Ethernet under a benign-only setting. The training and validation splits contain only benign traffic. The test split mixes benign traffic with multiple attack types.
Here, \textit{unsupervised} means that no attack samples or attack labels are used to optimize the model.
This is also a one-class anomaly-detection setting.
Unlike the semi-supervised IDS in~\cite{ShiblyHITK23}, X-SPUR does not train on labeled attacks.
X-SPUR learns through next-token prediction \cref{eq:clm_loss}.
It differs from AERO~\cite{Jeong2024AERO} in its scoring mechanism, not its supervision regime.
AERO scores distance to a learned prototype. X-SPUR instead uses token-level surprisal.

Detection is performed at the window level, and each sample is a sliding window of at most $w=10$ consecutive packets extracted from a flow. For evaluation, a test window is labeled anomalous if any of its packets is an attack packet, and its attack type is inherited from the packet-level annotation.

Protocol-wise calibration statistics and candidate thresholds are computed from benign validation data (\cref{subsec:calibration,subsec:decision_rule}). For benchmark reporting, we use test labels to select the best-F1 operating percentile. This selection does not update model weights or calibration statistics.

\subsection{Datasets and Packet-to-Text Representation}
\label{subsec:towids}

TOW-IDS~\cite{Han2023TOWIDS} comprises two packet-dump captures from an automotive Ethernet testbed plus an extended normal-driving capture. We follow the dataset split strategy discussed in AERO~\cite{Jeong2024AERO} (cf.\ Table~II therein), with the addition of a large benign capture containing ${\approx}7$M packets (the \emph{Train-Huge} split) for pretraining.
The test split contains five attack scenarios (CAN DoS, CAN Replay, AVTP Frame Injection, MAC Flooding, and PTP Sync Injection), summarized in \cref{tab:towids_attacks}.

\begin{table}
\caption{Attack types in TOW-IDS}
\label{tab:towids_attacks}
\centering
\footnotesize
\setlength{\tabcolsep}{2.5pt}
\begin{tabular}{@{}lrl@{}}
\toprule
\textbf{Attack Name} & \textbf{\# Windows} & \textbf{Description} \\
\midrule
CAN DoS             & 126,651 & Flood of CAN-over-UDP packets \\
CAN Replay          & 161,370 & Replayed CAN-over-UDP packets \\
AVTP Frame Injection & 61,894 & Injected AVTP video frames \\
MAC Flooding        & 9,000 & Random MAC address flooding \\
PTP Sync Injection  & 90,006 & Spoofed gPTP synchronization \\
\bottomrule
\end{tabular}
\end{table}

The benign traffic spans three protocol families with distinct statistical characteristics---UDP (CAN-over-UDP), AVTP (IEEE~1722 audio/video), and gPTP (IEEE~802.1AS)---whose markedly different score distributions later motivate protocol-aware score calibration. Minor protocol types (ARP, 0x88f5, and SRP) are excluded owing to their limited sample sizes.

To assess the transferability of the proposed methodology, we further evaluate X-SPUR on the automotive Ethernet (AE) portion of CarDS~\cite{Hellemans2025CarDS}. CarDS was captured from a 2020 commercial electric vehicle and contains 181M AE messages over 258 traces. Unlike TOW-IDS, its AE network is IPv6-based and VLAN-segmented. It carries general-purpose IP protocols such as TCP, UDP, ICMPv6, the Real-time Transport Protocol (RTP), and HTTP. We use the same model architecture and training procedure for its AE traffic. Benign traces form the train and validation splits. The test set contains AE pentesting traces, including Nmap scans with MAC/IP/VLAN spoofing, rear-view-camera RTP replacement, and an in-vehicle infotainment (IVI) REST-API crash (\cref{tab:cards}).

\begin{table}[t]
\caption{CarDS~\cite{Hellemans2025CarDS} automotive Ethernet dataset composition used for cross-dataset evaluation.}
\label{tab:cards}
\centering
\footnotesize
\setlength{\tabcolsep}{4pt}
\renewcommand{\arraystretch}{1.05}
\begin{tabular}{@{}llrr@{}}
\toprule
\textbf{Split} & \textbf{Class} & \textbf{\# Traces} & \textbf{\# Packets} \\
\midrule
Train-Huge & Benign & 38 & 14,276,579 \\
Train      & Benign & 27 &  7,239,662 \\
Validation & Benign &  6 &  4,120,212 \\
\midrule
\multirow{6}{*}{Test}
 & Benign                       &  5 &  2,916,705 \\
 & Nmap SYN scan (\texttt{-sS}) & 18 &  6,002,445 \\
 & Nmap ACK scan (\texttt{-sA}) &  5 &  1,774,426 \\
 & Nmap FIN scan (\texttt{-sF}) &  2 &    543,372 \\
 & RTP camera replacement       &  5 &  2,762,142 \\
 & IVI REST API crash           &  1 &    133,813 \\
\bottomrule
\end{tabular}
\end{table}

\begin{figure*}[t]
    \centering
    \includegraphics[width=1.0\textwidth]{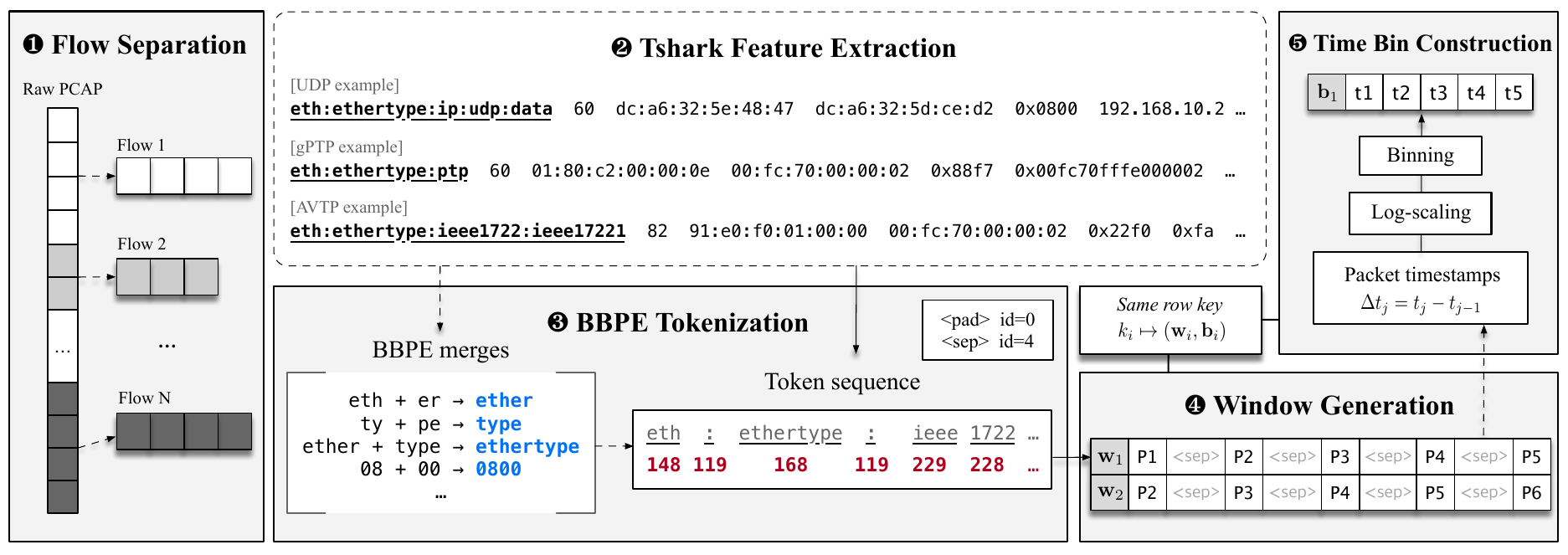}
    \caption{Overview of the X-SPUR preprocessing pipeline. Raw PCAP traffic is separated into flows, parsed with \texttt{tshark}, converted to BBPE token sequences, grouped into sliding windows, and aligned with log-scaled inter-packet time features.}
    \label{fig:preprocessing_example}
\end{figure*}

The preprocessing pipeline (\cref{fig:preprocessing_example}) separates raw PCAP files into per-flow captures keyed by source/destination MAC address, source/destination IP address, and transport-layer ports. Each flow is then parsed with \texttt{tshark} into a text representation spanning up to 180 protocol fields across the link, VLAN, and network/transport layers, together with protocol-specific fields from the IEEE~1722 (AVTP), IEC~61883, MPEG-TS, H.264, MPEG-PES, and PTP/gPTP layers.

\subsection{BBPE Tokenization and Window Construction}
\label{subsec:token_window}

To convert packet text into model input tokens, we train a BBPE tokenizer with vocabulary size $|\mathcal{V}| = 16{,}000$ on the Train-Huge split. The vocabulary reserves special tokens for padding (\texttt{<pad>}), unknown bytes, source/destination IP markers, packet boundaries (\texttt{<sep>}), and end of sequence. The resulting subword tokens preserve protocol-specific patterns while remaining flexible across heterogeneous field values. For example, the string \texttt{eth:ethertype:ip:udp:data} is tokenized as \texttt{[eth, :, ethertype, :, ip, :, udp, :, data]}. Tokenized flows are then segmented into sliding windows ($w=10$ packets, stride~1), and the packet token sequences within a window are concatenated with \texttt{<sep>} tokens at packet boundaries. Sequences are truncated to their first $L_{\max}$ tokens and padded if shorter.

\subsection{Time Representation and Window Labels}
\label{subsec:time_and_labels}

In addition to packet content, each window carries aligned inter-packet timing. Raw packet timestamps are converted into inter-packet time deltas $\Delta t_i = t_i - t_{i-1}$ and log-scaled as
\begin{equation}
s_i = \log_{10}(\Delta t_i + 10^{-7}).
\label{eq:log_time_section3}
\end{equation}
The values are clipped to $[-7,7]$ and stored using 1{,}000 uniform quantization bins. Their bin midpoints are fed as continuous scalars to the time-projection multilayer perceptron (MLP) in \cref{eq:time_mlp}, rather than through a learned per-bin embedding. Finally, each test window is assigned its binary label by mapping window indices back to the per-packet annotations, following the rule in \cref{subsec:task_setting}.

\section{Proposed X-SPUR Framework}
\label{sec:proposed_framework}

\begin{figure*}[t]
    \centering
    \includegraphics[width=1.0\textwidth]{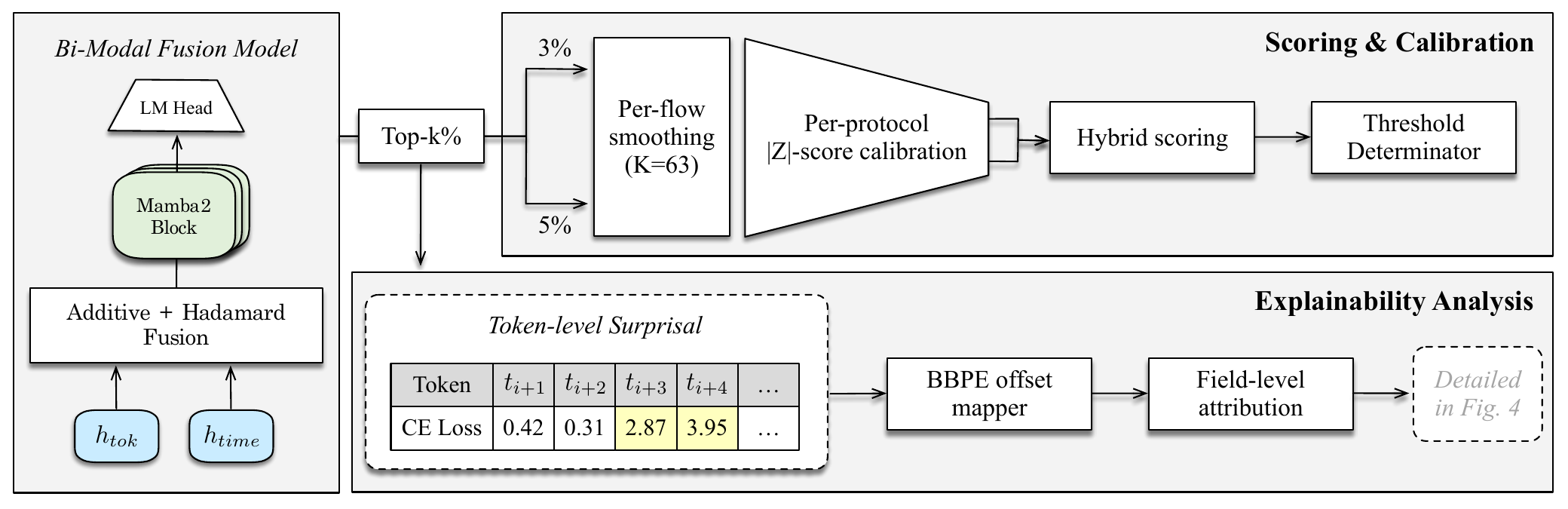}
    \caption{Overall framework of X-SPUR, from preprocessing and bimodal causal modeling to calibration and explainability.}
    \label{fig:framework_overview}
\end{figure*}

As illustrated in \cref{fig:framework_overview}, the X-SPUR framework operates in three stages. During training, a bimodal causal language model learns to predict the next token in benign traffic. During calibration, validation-set scores are computed at two top-$k$ levels, smoothed within each flow, and normalized per protocol. During detection, a test window is scored by token-level cross-entropy, converted into top-$k$ window scores, smoothed, calibrated, and thresholded. The same cross-entropy values provide field-level attribution.

\subsection{Mamba2-Based Causal Language Model}
\label{subsec:mamba_predictor}

The backbone of X-SPUR is a causal language model built on the Mamba2 state-space architecture~\cite{dao2024mamba2}. Its $O(L)$ sequence processing limits the growth in computational cost as the input context increases (\cref{sec:background}).

The base architecture embeds each token and its position through a token embedding $\mathbf{e}_{\mathrm{tok}}:\mathcal{V}\rightarrow\mathbb{R}^{d}$ and a learnable absolute positional embedding $\mathbf{e}_{\mathrm{pos}}$, applies dropout to their sum, and processes the sequence with a stack of $N$ pre-norm residual Mamba2 blocks,
\begin{equation}
\mathbf{x}^{(\ell+1)} = \mathbf{x}^{(\ell)} + \mathrm{Mamba2}\!\left(\mathrm{LN}\!\left(\mathbf{x}^{(\ell)}\right)\right),
\label{eq:mamba_residual}
\end{equation}
followed by a final layer normalization and a language-model head whose output weights are tied to the token embedding matrix.

Each Mamba2 block can be viewed as a selective state-space model: given an input sequence $x(t)$, the hidden state evolves as
\begin{equation}
h'(t) = A\,h(t) + B(t)\,x(t),
\qquad
y(t) = C(t)\,h(t),
\label{eq:ssm_continuous}
\end{equation}
where $A$ is a structured (diagonal) state-transition matrix and the projections $B(t)$ and $C(t)$ are input-dependent. This selectivity lets the model modulate its state by the current token context, which suits detecting unusual protocol-field transitions. For discrete sequences, the continuous model is discretized with a learned step size $\Delta$, yielding recurrent updates computed efficiently by a parallel scan.

\subsection{Bimodal Fusion of Payload and Inter-Packet Time}
\label{subsec:bimodal_fusion}

To incorporate temporal context beyond packet content, X-SPUR extends the base model with a bimodal fusion module that combines token, positional, and time embeddings through both additive fusion and an element-wise Hadamard interaction:
\begin{equation}
\begin{aligned}
\mathbf{x}_i & =
\mathbf{e}_{\mathrm{tok}}(x_i)+\mathbf{e}_{\mathrm{pos}}(i)+\mathbf{e}_{\mathrm{time}}(s_i) \\
&\quad + \alpha \left(\mathbf{e}_{\mathrm{tok+pos}} \odot \mathbf{e}_{\mathrm{time}}(s_i)\right),
\end{aligned}
\label{eq:bimodal_fusion}
\end{equation}
where $\alpha=1.0$ is a fixed interaction strength, $\odot$ denotes element-wise multiplication, $\mathbf{e}_{\mathrm{time}}(s_i)\in\mathbb{R}^d$ is a learned embedding of the log-scaled inter-packet delta $s_i$ from \cref{eq:log_time_section3}, and $\mathbf{e}_{\mathrm{tok+pos}}$ abbreviates the sum $\mathbf{e}_{\mathrm{tok}}(x_i)+\mathbf{e}_{\mathrm{pos}}(i)$.

The time embedding itself is produced by a two-layer MLP:
\begin{equation}
\mathbf{e}_{\mathrm{time}}(s)
=
W_2\,\mathrm{GELU}\!\left(\mathrm{LN}(W_1 s + b_1)\right)+b_2,
\label{eq:time_mlp}
\end{equation}
where $W_1 \in \mathbb{R}^{d_{\mathrm{proj}}\times 1}$, $W_2 \in \mathbb{R}^{d\times d_{\mathrm{proj}}}$, $d_{\mathrm{proj}}=64$, and dropout with $p=0.1$ is applied between the two linear layers.

Because BBPE yields a variable number of subword tokens per packet while time deltas are per packet, $s_j$ is replicated across all token positions of packet $j$.

The model uses dimension $d=768$ with $N=24$ layers and maximum sequence length $L_{\max}=256$. We use head dimension 128, state dimension $d_{\mathrm{state}}=16$, convolution width $d_{\mathrm{conv}}=4$, and dropout 0.1. The model has ${\approx}98.5$M parameters.

\subsection{Benign-Only Training Objective}
\label{subsec:training_objective}

On TOW-IDS, training is benign-only and two-phase: pretraining on the Train-Huge split (${\approx}7$M benign windows), then fine-tuning on the smaller Train split (${\approx}228$k windows) to stabilize learned patterns and match the target distribution. We pretrain for 10 epochs and fine-tune for 3 epochs with batch size 32. Both phases use the same causal language-modeling objective.

For a model input window $X=(x_{1:L},s_{1:L})$, comprising a token sequence and its aligned time features, let $\mathcal{I}_X=\{j\in\{2,\ldots,L\}:x_j\neq\texttt{<pad>}\}$ denote the valid target-token positions. The training loss is the standard next-token cross-entropy:
\begin{equation}
\mathcal{L}_{\mathrm{CLM}}
=
-\frac{1}{|\mathcal{I}_X|}\sum_{j\in\mathcal{I}_X} \log P(x_j\mid x_{<j},s_{<j};\Theta),
\label{eq:clm_loss}
\end{equation}
where $\Theta$ denotes the model parameters and \texttt{<pad>} positions are excluded from the sum.
\Cref{alg:xspur_train} summarizes the offline training and calibration procedure.

\subsection{Window-Level Anomaly Scoring via Token Surprisal}
\label{subsec:anomaly_scoring}

\Cref{alg:xspur_infer} summarizes detection and attribution.
At inference time, a test window $X=(x_{1:L},s_{1:L})$ is scored using token-level causal language-model cross-entropy. For each $j\in\mathcal{I}_X$,
\begin{equation}
\mathrm{CE}_j = -\log P(x_j\mid x_{<j},s_{<j};\Theta),
\label{eq:token_ce}
\end{equation}
where positions corresponding to \texttt{<pad>} are masked out. The value $\mathrm{CE}_j$ measures the surprisal of target token $x_j$ under the benign model. Unusually high values indicate deviations from learned protocol behavior.

Rather than averaging cross-entropy across the full window, X-SPUR computes a top-$k$\% anomaly score:
\begin{equation}
S_k(X)
=
\frac{1}{|T_k|}
\sum_{j\in T_k}\mathrm{CE}_j,
\label{eq:topk_score}
\end{equation}
where $T_k$ contains the $\max\{1,\lfloor k|\mathcal{I}_X|\rfloor\}$ positions in $\mathcal{I}_X$ with the highest cross-entropy values. Many attacks affect only a small subset of packet fields, producing sparse surprisal spikes that whole-window averaging would dilute. Top-$k$ scoring preserves their contribution. We use two settings, $k=5\%$ and $k=3\%$.

\subsection{Flow-Level Smoothing and Protocol-Aware Calibration}
\label{subsec:calibration}

Since neighboring windows within the same flow overlap in nine out of ten packets, their raw anomaly scores are highly correlated. To reduce transient noise, we apply moving-average smoothing to both top-$k$ scores within each flow, using kernel size $K=63$ (selected in \cref{sec:experiments}):
\begin{equation}
\tilde{S}_i
=
\frac{1}{K}
\sum_{j=i-\lfloor K/2 \rfloor}^{i+\lfloor K/2 \rfloor}
S_j.
\label{eq:smoothing}
\end{equation}
At flow boundaries, the nearest endpoint score is replicated. Flows with at most $K$ windows are left unsmoothed.

Because benign score levels differ sharply across protocol families (gPTP windows sit far above UDP), X-SPUR normalizes per protocol. For each protocol $p$, identified from the flow's \texttt{frame.protocols} field, and each top-$k$ fraction $k$, the mean $\mu_{p,k}$ and standard deviation $\sigma_{p,k}$ of the smoothed validation scores are computed. A test score is then normalized as
\begin{equation}
z_k = \frac{\tilde{S}_k-\mu_{p,k}}{\max(\sigma_{p,k},\epsilon)},
\label{eq:zscore}
\end{equation}
where $\epsilon=10^{-6}$ ensures numerical stability. If protocol $p$ is absent from the benign validation split, the corresponding global validation mean and standard deviation are used as a fallback. On CarDS, this fallback also applies to protocols with fewer than 200 benign validation windows. Since both unusually high and unusually low scores may indicate abnormal behavior, the absolute value $|z_k|$ enters the final calibration stage.

Empirically, the two granularities are complementary: the top-5\% view is essential for PTP Sync Injection---its moderately distributed evidence all but vanishes from the narrower top-3\% token set---whereas the top-3\% view slightly sharpens sparse-spike attacks whose evidence concentrates in a few fields. The final hybrid calibrated score is therefore defined as
\begin{equation}
S_{\mathrm{hybrid}} = \max\left(|z_{0.05}|,\;|z_{0.03}|\right),
\label{eq:hybrid_score}
\end{equation}
where $z_{0.05}$ and $z_{0.03}$ denote the calibrated top-5\% and top-3\% scores, respectively. Whichever granularity produces the more extreme deviation drives detection.

\subsection{Decision Rule and Threshold Selection}
\label{subsec:decision_rule}

For a fixed operating percentile $\rho$, the decision threshold $\tau$ is computed from the calibrated benign-only validation scores.
A window is flagged as anomalous if and only if $S_{\mathrm{hybrid}} \ge \tau$.
In practice, a domain expert should select the operating percentile based on the deployment's tolerance for false alarms.

Following the evaluation protocol of AERO~\cite{Jeong2024AERO}, we sweep $\rho$ from the 90th to the 99.99th percentile.
For benchmark reporting, we select the percentile with the highest test-set F1-score.
The best-F1 operating point is attained at the 99.94th percentile ($\tau = 3.9122$ on the hybrid score).
It yields a test precision of $0.9785$ and recall of $0.9775$.
This reporting choice does not update model weights or calibration statistics.

\subsection{Intrinsic Explainability via Field Attribution}
\label{subsec:explainability_method}

Because the window-level score is derived from token-level cross-entropy, field-level attribution requires no additional post-hoc explanation model. It proceeds directly from the scored window. The BBPE tokenizer's offset mapping locates each token's character span in the original packet text. The span is mapped to its tab-separated field index and field name in the \texttt{tshark} schema. Token-level cross-entropy values are then aggregated within each field:
\begin{equation}
\mathrm{CE}_f
=
\frac{1}{|T_f|}
\sum_{j\in T_f}\mathrm{CE}_j,
\label{eq:field_ce}
\end{equation}
where $T_f$ is the set of valid target-token positions $j$ whose token $x_j$ belongs to field $f$.

This token-to-field mapping traces anomaly scores back to concrete protocol fields and reveals distinct attack signatures localized to the specific fields each intrusion disturbs. Two properties delimit what such an explanation means. First, it is directly grounded in the model score. Equation~\cref{eq:field_ce} aggregates field-associated values from the same token-level cross-entropies used to form the window score. These values precede the top-$k$ pooling, smoothing, and calibration of \cref{subsec:anomaly_scoring} and \cref{subsec:calibration}. Unlike a separately fitted surrogate, it therefore reports local model evidence rather than an approximation from another model. Second, surprisal marks deviation from learned benign regularity, not maliciousness itself: attribution answers \emph{which fields deviate}, and \cref{subsec:fp_analysis} shows the same readout diagnosing residual false positives.

\begin{algorithm}[t]
\small
\DontPrintSemicolon
\KwInput{\\
Benign PCAP splits $\mathbb{P}_{\mathrm{huge}}$, $\mathbb{P}_{\mathrm{tr}}$, and $\mathbb{P}_v$; fixed \texttt{tshark} field schema\\
Top-$k$ fractions $\mathcal{R}=\{0.03,0.05\}$; fixed operating percentile $\rho$
}
\KwOutput{\\
Frozen BBPE tokenizer $\mathcal{B}$; trained parameters $\Theta$\\
Statistics $\{(\mu_{p,k},\sigma_{p,k})\}_{p,k}$ and global fallback $\{(\mu_{*,k},\sigma_{*,k})\}_k$\\
Threshold $\tau$
}
\tcp{Step 1---Data preparation (\cref{subsec:token_window} and \cref{subsec:time_and_labels})}
For each $d\in\{\mathrm{huge},\mathrm{tr},v\}$, split $\mathbb{P}_d$ into ordered flows and serialize the fixed packet fields with \texttt{tshark} $\rightarrow \mathbb{R}_d$.\;
Train $\mathcal{B}$ ($|\mathcal{V}|=16{,}000$) on $\mathbb{R}_{\mathrm{huge}}$ only.\;
For each $d\in\{\mathrm{huge},\mathrm{tr},v\}$, BBPE-encode $\mathbb{R}_d$, insert \texttt{<sep>}, align log-scaled time deltas, and form stride-1 windows ($w=10$, $L_{\max}=256$) $\rightarrow\mathbb{S}_d$.\;
\tcp{Step 2---Benign-only model fitting (\cref{subsec:training_objective})}
Initialize $\Theta$.\;
\ForEach{$(\mathbb{S},E)\in[(\mathbb{S}_{\mathrm{huge}},10),(\mathbb{S}_{\mathrm{tr}},3)]$ in order}{
  \For{epoch $\leftarrow 1$ \KwTo $E$}{
    \For{mini-batch $X\subset\mathbb{S}$}{
      Fuse the token, position, and time embeddings; forward once through the Mamba2 model.\;
      Compute $\mathcal{L}_{\mathrm{CLM}}$ \cref{eq:clm_loss} and update $\Theta$ with AdamW.\;
    }
  }
}
\tcp{Step 3---Benign validation calibration (\cref{subsec:calibration})}
\For{each ordered validation window $X_i\in\mathbb{S}_v$}{
  Forward $X_i$ once and retain its valid target-token surprisals $\{\mathrm{CE}_{i,j}\}_j$.\;
  \For{$k\in\mathcal{R}$}{Compute $S_{i,k}$ from the largest $k$ fraction of $\{\mathrm{CE}_{i,j}\}_j$ \cref{eq:topk_score}.\;}
}
\For{$k\in\mathcal{R}$}{
  Smooth $\{S_{i,k}\}_i$ separately within each flow \cref{eq:smoothing} $\rightarrow\{\tilde{S}_{i,k}\}_i$.\;
  \For{each protocol $p$ with sufficient validation windows (\cref{subsec:calibration})}{
    $(\mu_{p,k},\sigma_{p,k})\leftarrow$ mean and standard deviation of $\tilde{S}_{i,k}$ for windows of protocol $p$.\;
  }
  $(\mu_{*,k},\sigma_{*,k})\leftarrow$ global mean and standard deviation of $\tilde{S}_{i,k}$.\;
}
\tcp{Step 4---Threshold (\cref{subsec:decision_rule})}
For every $X_i\in\mathbb{S}_v$, compute $z_{i,k}$ with its protocol statistics or the global fallback \cref{eq:zscore} and set $S_{i,\mathrm{hybrid}}=\max_{k\in\mathcal{R}}|z_{i,k}|$.\;
$\tau\leftarrow\operatorname{Percentile}_{\rho}(\{S_{i,\mathrm{hybrid}}\}_i)$.\;
\caption{Offline Training and Calibration}
\label{alg:xspur_train}
\end{algorithm}

\begin{algorithm}[t]
\small
\DontPrintSemicolon
\KwInput{\\
Raw test flow $F$ of protocol $p$; frozen $\mathcal{B}$ and field schema\\
Trained $\Theta$; calibration and fallback statistics from \Cref{alg:xspur_train}; $\mathcal{R}=\{0.03,0.05\}$; threshold $\tau$
}
\KwOutput{\\
Window decisions $y_i\in\{0,1\}$, one per resulting window\\
For every $y_i=1$, a ranked field attribution
}
Apply the frozen preprocessing of \Cref{alg:xspur_train} to $F$, retaining text and BBPE offsets $\rightarrow(X_1,\ldots,X_n)$.\;
\For{$i\leftarrow1$ \KwTo $n$}{
  Forward $X_i$ once; retain $\{\mathrm{CE}_{i,j}\}_j$ \cref{eq:token_ce}.\;
  \For{$k\in\mathcal{R}$}{Compute $S_{i,k}$ from the same retained surprisals \cref{eq:topk_score}.\;}
}
\For{$k\in\mathcal{R}$}{Apply the centered $K=63$ smoother within $F$ \cref{eq:smoothing} $\rightarrow\{\tilde{S}_{i,k}\}_{i=1}^{n}$.\;}
For each $k\in\mathcal{R}$, set $(\bar\mu_k,\bar\sigma_k)$ to $(\mu_{p,k},\sigma_{p,k})$ if available, otherwise to $(\mu_{*,k},\sigma_{*,k})$.\;
\For{$i\leftarrow1$ \KwTo $n$}{
  $z_{i,k}\leftarrow(\tilde{S}_{i,k}-\bar\mu_k)/\max(\bar\sigma_k,\epsilon)$ for each $k\in\mathcal{R}$.\;
  $S_{i,\mathrm{hybrid}}\leftarrow\max(|z_{i,0.03}|,|z_{i,0.05}|)$; $y_i\leftarrow1$ if $S_{i,\mathrm{hybrid}}\geq\tau$, else $y_i\leftarrow0$.\;
  \If{$y_i=1$}{
    Map the retained surprisals to \texttt{tshark} fields via BBPE offsets; rank fields by $\mathrm{CE}_f$ \cref{eq:field_ce}.\;
  }
}
\caption{Detection and Attribution}
\label{alg:xspur_infer}
\end{algorithm}

\section{Experimental Results}
\label{sec:experiments}
\subsection{Experimental Setup}
\label{subsec:experimental_setup}
Unless stated otherwise, experiments use benign-only pretraining and fine-tuning. They use the TOW-IDS test split and the same scoring procedure. The ablations in Sections V-B--V-D vary only the settings under study around the default configuration ($w=10$, sequence length 256, $K=63$, $k\in\{3\%, 5\%\}$). We report threshold-independent AUC. For threshold-dependent metrics, we report the best F1-score over the stated sweep of validation-derived percentile thresholds. Precision and recall are reported at that operating point. We also report per-attack false-negative rates (FNRs) and per-protocol false-positive rates (FPRs).

\subsection{Context and Smoothing-Width Selection}
\label{subsec:context_selection}
\begin{figure}[t]
    \centering
    \includegraphics[width=1.0\columnwidth]{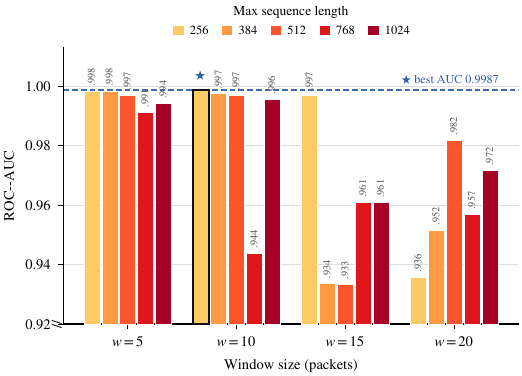}
    \caption{Context sweep over window size $w$ and maximum sequence length (AUC), grouped by $w$ and shaded by sequence length; the dashed line marks the best AUC and the starred bar ($w=10$, sequence length 256) the selected operating point.}
    \label{fig:context_sweep}
\end{figure}

\cref{fig:context_sweep} reports the context sweep over window size $w$ and maximum sequence length. Shorter windows generally give higher AUC. Larger windows show no consistent gain and greater variation across sequence lengths. We therefore adopt $w=10$ with sequence length 256, the best operating point ($\mathrm{AUC}=0.9987$) and the default for all subsequent analyses.

To select the smoothing width, we sweep $K\in\{7,15,31,63,127\}$ under this context setting. AUC rises steadily with the smoothing window, from 0.9901 at $K=7$ to its maximum 0.9987 at $K=63$, then falls to 0.9924 at $K=127$. We therefore fix $K=63$. Smoothing is decisive for the hardest attack: at the same validation-percentile threshold, unsmoothed scores miss 97\% of AVTP Frame Injection windows versus 16\% after smoothing, because the weak per-window evidence of injected frames becomes detectable only when accumulated across a contiguous burst.

\BfPara{Run-to-run results} We repeated the TOW-IDS experiment in five additional runs with different random seeds. The mean AUC was 0.9939, with a sample standard deviation of 0.0041. The AUC ranged from 0.9880 to 0.9985. The AUC of 0.9987 in Tables~\ref{tab:comparison_prior_art} and~\ref{tab:backbone_comparison} is from the original experiment (seed 42).
These results indicate consistently high detection performance across repeated runs.

\subsection{Time-Fusion Ablation}
To isolate the contribution of inter-packet timing, we retrain the model with the time channel removed (\emph{payload only}) using the same context settings. Discarding the time modality lowers the AUC from 0.9987 to 0.9962 and the F1-score from 0.9780 to 0.9663, with a consistent drop across precision and recall. This is expected in automotive Ethernet, where replayed, flooded, and spoofed-synchronization attacks perturb inter-packet timing as much as payload---deviations the time-aware embedding surfaces directly but a payload-only model can only infer from content.

\subsection{Calibration Ablation}
\begin{table}[t]
\caption{Calibration ablation across the two top-$k$ views; ``Prot.-cal.'' denotes per-protocol $Z$-score normalization.}
\label{tab:calibration}
\centering
\footnotesize
\setlength{\tabcolsep}{3.6pt}
\renewcommand{\arraystretch}{1.0}
\begin{adjustbox}{max width=\linewidth}
\begin{tabular}{@{}llrr@{}}
\toprule
Top-$k$ & Stage & AUC & F1-score \\
\midrule
3\% & Raw        & 0.9838 & 0.9296 \\
3\% & Smooth.    & 0.9942 & 0.9518 \\
3\% & Prot.-cal. & 0.9958 & 0.9604 \\
\midrule
5\% & Raw        & 0.9763 & 0.9087 \\
5\% & Smooth.    & 0.9907 & 0.9411 \\
5\% & Prot.-cal. & 0.9986 & 0.9749 \\
\midrule
\multicolumn{2}{@{}l}{Hybrid $\max(|z_{0.05}|,|z_{0.03}|)$} & \textbf{0.9987} & \textbf{0.9780} \\
\bottomrule
\end{tabular}
\end{adjustbox}
\end{table}

\cref{tab:calibration} shows a consistent progression: raw top-$k$ scores separate benign and malicious traffic reasonably well, flow-level smoothing pushes both views above 0.99 AUC, per-protocol $Z$-score normalization further stabilizes the distributions, and the hybrid score fuses the two calibrated views for the best operating point (AUC 0.9987).

\subsection{Per-Attack Performance}
\cref{tab:per_attack_ws10} reports the attack-wise FNR of X-SPUR versus AERO~\cite{Jeong2024AERO}, together with each attack's mean hybrid anomaly score (\cref{eq:hybrid_score}). CAN DoS, CAN Replay, and MAC Flooding are detected essentially perfectly. PTP Sync Injection has well above 99\% recall, only marginally behind AERO. AVTP Frame Injection remains hardest by a wide margin. Its evidence is confined to a few IEC\,61883 stream fields already unpredictable in benign traffic (\cref{sec:explainability_results}).

\begin{table}[t]
\caption{Per-attack false-negative rate (FNR) versus AERO~\cite{Jeong2024AERO}, with mean hybrid anomaly score.}
\label{tab:per_attack_ws10}
\centering
\footnotesize
\setlength{\tabcolsep}{4.0pt}
\renewcommand{\arraystretch}{1.0}
\begin{tabular}{l r r r}
\toprule
\textbf{Attack} & \makecell{\textbf{X-SPUR}} & \makecell{\textbf{AERO}} & \makecell{\textbf{Mean} $S_{\mathrm{hybrid}}$} \\
\midrule
CAN DoS         & \textbf{0.00\%} & 0.86\%          &  8.621 \\
CAN Replay      & \textbf{0.14\%} & 4.25\%          &  9.607 \\
AVTP Frame Inj.  & 15.77\%         & \textbf{2.43\%} &  5.315 \\
MAC Flooding    & \textbf{0.00\%} & 3.30\%          & 21.157 \\
PTP Sync Inj.   & 0.13\%          & \textbf{0.04\%} &  5.495 \\
\bottomrule
\end{tabular}
\end{table}

\subsection{Field-Level Explainability Analysis}
\label{sec:explainability_results}
\begin{figure*}[p]
    \centering
    \includegraphics[trim=24 26 24 24, clip, width=\textwidth]{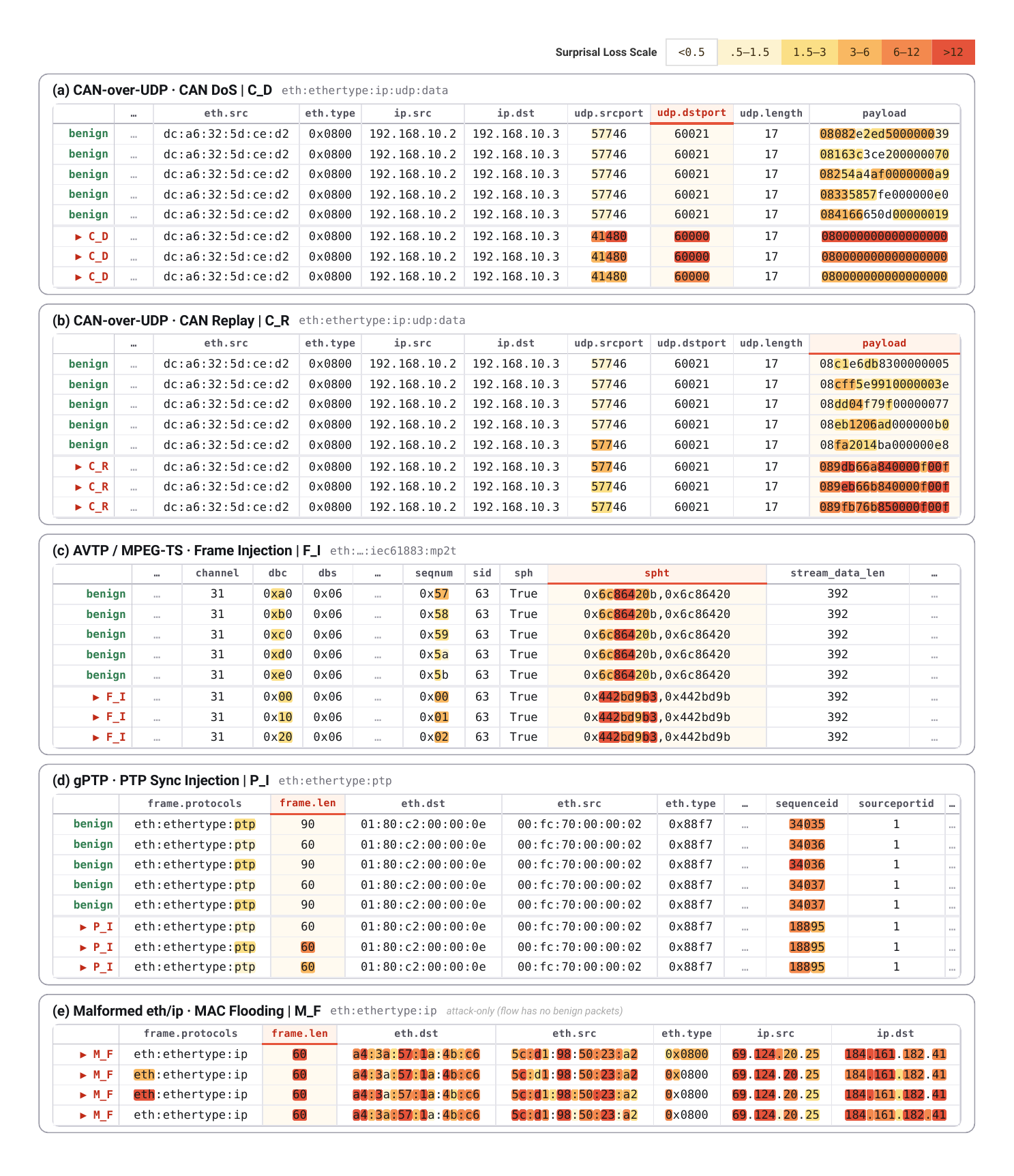}
    \caption{Intrinsic field-level explainability of X-SPUR on TOW-IDS. Panels (a)--(e) show representative test examples for the five attack classes. Columns are \texttt{tshark} fields. Cells hold the raw tokens shaded by mean per-token cross-entropy (surprisal). The most-discriminative field is highlighted. Structurally fixed fields stay cold while the attacked field ignites: \texttt{udp.dstport} for CAN DoS~(a) and the replayed \texttt{payload} for CAN Replay~(b). PTP Sync Injection~(d) shows the collapsed Sync/Follow-Up cadence in \texttt{frame.len}. MAC Flooding~(e) shows a uniform elevation across every header. AVTP Frame Injection~(c) is hardest: its valid frame only resets the already-warm \texttt{iec61883.spht} to an equally unpredictable value, so surprisal barely rises. Ellipses~(\ldots) denote cold, low-surprisal fields omitted to fit the figure width.}
    \label{fig:protocol_explainability}
\end{figure*}

\cref{fig:protocol_explainability} shows X-SPUR's intrinsic field-level attribution (\cref{subsec:explainability_method}) in five panels, (a)--(e), using TOW-IDS test examples.
All panels use the model and settings in \cref{subsec:experimental_setup}, with $w=10$ and $L_{\max}=256$.
Each panel has columns for the \texttt{tshark} fields.
Each token is shaded by its cross-entropy \cref{eq:token_ce} averaged over the sliding windows scoring it.
Attribution reuses the token scores computed for detection.
Reading a field's column top-to-bottom exposes deviations from learned benign behavior, and the most-discriminative field is highlighted.
The panels reveal how differently each attack manifests.
Some ignite only one or two fields while the rest stay cold.
Others warm nearly every header field at once.
Still others surface only as a broken cadence in a single structural field.
Because attribution operates on the decoded \texttt{tshark} schema, the panels name the responsible fields directly.

The field-level patterns are consistent with the per-attack results in \cref{tab:per_attack_ws10}.
CAN DoS and CAN Replay show clear field-level changes and near-zero false-negative rates.
AVTP Frame Injection shows weaker contrast and the highest false-negative rate.

\BfPara{CAN-over-UDP (\cref{fig:protocol_explainability}(a,b))} Both CAN attacks ignite only transport and payload fields and leave every Ethernet/IP header cold. CAN DoS~(a) surprises the model at \texttt{udp.dstport}---the spoofed attacker port \texttt{60000}, never seen in benign traffic, reaches a cross-entropy near~12---together with the injected \texttt{payload} and \texttt{udp.srcport}. CAN Replay~(b) instead lights only \texttt{payload}: the replayed frames reuse valid headers and ports, so the sole anomaly is the stale byte sequence. Fewer than 1\% of tokens are extreme, and this tight localization yields near-zero false negatives.

\BfPara{AVTP Frame Injection (\cref{fig:protocol_explainability}(c))} The injected frames are structurally valid MPEG-TS content, so the video payload stays cold; the injection instead resets the IEC\,61883 stream counters (\texttt{iec61883.seqnum}, \texttt{iec61883.dbc}) and the media timestamp \texttt{iec61883.spht}. Yet \texttt{spht} carries an arbitrary media-clock timestamp already highly surprising in benign traffic, so the injection lifts it only marginally and no field turns cleanly from cold to warm. This absence of a sharp field-level signal is the visible reason AVTP Frame Injection keeps the highest false-negative rate (\cref{tab:per_attack_ws10}); detection instead rests on the aggregate window score.

\BfPara{PTP Sync Injection (\cref{fig:protocol_explainability}(d))} Benign gPTP alternates Sync and Follow-Up messages of 60 and 90 bytes; the Sync flood collapses this cadence into a run of 60-byte Syncs, so \texttt{frame.len} ignites, while the sequence-id counter stays high in both benign and attack and is not the discriminator. Surprisal reflects spoofed synchronization semantics rather than corrupted payload, and detection stays at 99.9\%.

\BfPara{MAC Flooding (\cref{fig:protocol_explainability}(e))} At the opposite extreme, malformed bare-IP frames whose random addresses never occur in benign traffic warm the entire header stack at once---the flood violates the learned packet structure as a whole, and recall is perfect.

Overall, apart from MAC Flooding the evidence stays confined to the few fields each attack actually perturbs rather than spreading across generic Ethernet or IP headers, and surprisal magnitude and locality together predict difficulty. Recall is 99.6\% at the first and last anomalous windows of each flow, compared with 97.7\% for interior windows.

\subsection{Per-Protocol False-Positive Analysis}
\label{subsec:fp_analysis}
\begin{table}[t]
\caption{Per-protocol false-positive rates.}
\label{tab:per_proto_fp_ws10}
\centering
\footnotesize
\setlength{\tabcolsep}{6pt}
\renewcommand{\arraystretch}{1.0}
\begin{tabular}{l r r r}
\toprule
\textbf{Protocol} & \textbf{FP} & \textbf{Total benign} & \textbf{FPR} \\
\midrule
AVTP  & 5,932 & 343,991 & 1.72\% \\
UDP   & 3,663 & 842,345 & 0.43\% \\
gPTP  & 51 & 11,448 & 0.45\% \\
\midrule
\textbf{Total} & \textbf{9,646} & \textbf{1,197,784} & \textbf{0.81\%} \\
\bottomrule
\end{tabular}
\end{table}

\cref{tab:per_proto_fp_ws10} splits the false positives by protocol (aggregate 0.81\%). The gPTP figure is the clearest sign that per-protocol $Z$-score calibration removes protocol-level bias rather than relocating a global threshold: gPTP carries one of the highest benign baseline surprisals yet ends at an FPR comparable to UDP.

The residual errors are localized, not diffuse: two AVTP MP2T streams alone produce 3,977 and 1,892 false-positive windows, with a few UDP flows adding most of the remainder. Crucially, the fields that trigger them are the same IEC\,61883/MPEG-TS stream fields, led by \texttt{iec61883.spht}, that carry the AVTP Frame Injection signal in \cref{fig:protocol_explainability}(c). The false positives and the hardest attack thus share one interpretable cause---the model's heightened sensitivity to the IEC\,61883/MPEG-TS region---which token-to-field attribution makes directly visible where a window-level score could not. The coupling is quantitative: missed AVTP Frame Injection windows are near-misses (median $z=3.4$ against $\tau=3.91$), but roughly twice as many benign AVTP windows occupy the same band, so recovering them by lowering $\tau$ would cost about two false positives per miss.

\subsection{Cross-Dataset Evaluation on CarDS}
\label{sec:cards_eval}
\begin{table}[t]
\caption{Cross-dataset detection on CarDS; AERO F1 at a matched 0.5\% benign-FP budget. The GRU and AERO AUCs coincide at this precision.}
\label{tab:cards_baselines}
\centering
\footnotesize
\setlength{\tabcolsep}{6pt}
\renewcommand{\arraystretch}{1.0}
\begin{tabular}{l cccc}
\toprule
\textbf{Metric} & \textbf{X-SPUR} & CNN & GRU & AERO~\cite{Jeong2024AERO} \\
\midrule
AUC & \textbf{0.9923} & 0.6030 & 0.6239 & 0.6239 \\
F1  & \textbf{0.9842} & 0.3577 & 0.3167 & 0.1074 \\
\bottomrule
\end{tabular}
\end{table}

To test the transferability of the proposed methodology beyond TOW-IDS, we train a separate X-SPUR model from scratch on CarDS~\cite{Hellemans2025CarDS}.
We use the 38-trace Train-Huge corpus to broaden benign-traffic coverage during pretraining.
Fine-tuning uses the 27-trace Train split (\cref{tab:cards}).
We retain the model architecture, pretrain-to-finetune procedure, and training hyperparameters used for TOW-IDS (\cref{subsec:training_objective}). We also fit the BBPE tokenizer on the Train-Huge split listed in \cref{tab:cards}. Calibration statistics and candidate thresholds are derived from its benign validation split.
Despite a markedly different protocol stack (IPv6-based TCP/UDP with RTP and HTTP services rather than AVTP/gPTP) and attack taxonomy (network-reconnaissance scans, RTP stream replacement, and an infotainment service crash), X-SPUR retains strong performance (AUC 0.9923, F1-score 0.9842). The baselines do not: CNN and GRU causal language models trained on the same tokenized input collapse to AUC 0.6030 and 0.6239 (\cref{tab:cards_baselines}), and an architecture-verified reimplementation of AERO, trained on the same CarDS benign splits and scored on the same test captures, peaks at AUC 0.6239 across four trained configurations---essentially blind to the network scans (recall ${\le}8\%$ at a 0.5\% false-positive budget) and only partially sensitive to RTP replacement.
Without changing the model architecture or training hyperparameters, X-SPUR achieves superior performance in this new environment.

Per attack family, RTP camera replacement is detected almost perfectly (FNR 0.03\%). The dominant Nmap SYN/ACK scans yield FNRs of 1.4--2.6\%. The smallest classes remain the hardest. Their FNRs are 7.3\% for FIN scans and 9.0\% for the single IVI REST-API-crash trace. The same per-token attribution pattern appears on CarDS. Surprisal concentrates on a small set of protocol fields. These include TCP ports and flags for the scans, frame length for RTP replacement, and crafted TCP connection fields for the REST crash. Surprisal magnitudes are roughly $40\times$ lower than on TOW-IDS. Per-protocol calibration accommodates this score-scale difference.

\subsection{Comparison with Prior Art}
\label{sec:comparison}
\cref{tab:comparison_prior_art} compares X-SPUR with representative automotive Ethernet baselines on TOW-IDS. These include reconstruction-based autoencoders, a semi-supervised classifier~\cite{ShiblyHITK23}, and AERO~\cite{Jeong2024AERO}. X-SPUR achieves an AUC of 0.9987. AERO reports 0.9969. This difference is marginal and does not establish statistical superiority. X-SPUR also uses a larger parameter budget (98.5M versus 289k). AERO remains better on two attack families (\cref{tab:per_attack_ws10}).

The first distinction is intrinsic field-level attribution. AERO's prototype-distance score does not natively map back to raw protocol fields. Its outlier score is the squared distance between a compressed 16-dimensional representation and a single learned prototype. Obtaining comparable field-level explanations would therefore require an additional attribution mechanism outside the detector. X-SPUR's field attribution instead re-aggregates the same per-token surprisal used for detection (\cref{subsec:explainability_method}). It averages the same token-level evidence within each field. This coupling is also reflected empirically in \cref{fig:protocol_explainability} and \cref{tab:per_attack_ws10}. AVTP Frame Injection shows weak field-level contrast between benign and attack traffic. It remains the hardest class (\cref{sec:explainability_results}).

The second distinction is the absence of protocol-specific handcrafted feature design. AERO's own feature extractor requires an expert to hand-select payload byte-offset parameters $(j,n)$ for each IVN~\cite{Jeong2024AERO}. Its AERO-FG2 ablation, using $(j,n){=}(0,434)$, shows that this choice materially affects detection performance. X-SPUR instead operates on a byte-level token representation that requires no protocol-specific feature engineering, avoiding the deployment-specific redesign required by AERO's feature extractor. 
The CarDS evaluation provides an empirical test of this property. We use the same model architecture and training hyperparameters on its markedly different protocol stack (\cref{subsec:experimental_setup}). A separate X-SPUR model is trained from scratch using CarDS benign data, without dataset-specific feature redesign or a search over training hyperparameters (\cref{sec:cards_eval}).

\begin{table}[t]
  \caption{Anomaly detection performance comparison on TOW-IDS
  (baseline results as reported in AERO~\cite{Jeong2024AERO}).}
  \label{tab:comparison_prior_art}

  \centering
  \footnotesize

  \begin{adjustbox}{width=\linewidth,center}
  \setlength{\tabcolsep}{2.5pt}

  \begin{tabular}{llrr}
    \toprule
    Method & Feature & \# Params & AUC \\
    \midrule
    Conv. autoencoder & Payload ($60 \times 60$) & 2,661k & 0.7904 \\
    Conv. autoencoder & Payload ($60 \times 60$), byte changes & 2,661k & 0.8103 \\
    Conv. autoencoder & Payload ($434 \times 434$) & 4,122k & 0.9560 \\
    Vanilla autoencoder & Time interval (mean, std., skew.) & 1,336k & 0.9648 \\
    Shibly~\textit{et al.}~\cite{ShiblyHITK23} & Payload (10\% labeled) & 112k & 0.9730 \\
    AERO~\cite{Jeong2024AERO} & Three multimodal features & 289k & 0.9969 \\
    \textbf{X-SPUR} & Raw packet tokens (BBPE) & 98.5M & \textbf{0.9987} \\
    \bottomrule
  \end{tabular}

  \end{adjustbox}
\end{table}

\subsection{Backbone Comparison}
\label{subsec:backbone_comparison}
\cref{tab:backbone_comparison} compares Mamba2 with Transformer, LSTM, and dilated TCN variants.
All models use the same TOW-IDS splits and token--time fusion, with $w=10$ and $L_{\max}=256$.
They share the dual top-$k$ scoring, $K=63$ smoothing, and per-protocol calibration.
The total parameter count of each alternative differs from Mamba2's by less than 1\%.
The Transformer also achieves a high AUC of 0.9944.
Mamba2 achieves the highest AUC of 0.9987, supporting its use as the X-SPUR backbone in this comparison.

\begin{table}[t]
\caption{Backbone comparison on TOW-IDS. Parameter counts are for the complete models.}
\label{tab:backbone_comparison}
\centering
\footnotesize
\setlength{\tabcolsep}{6pt}
\renewcommand{\arraystretch}{1.0}
\begin{tabular}{l r r}
\toprule
\textbf{Backbone} & \textbf{\# Params} & \textbf{AUC} \\
\midrule
\textbf{Mamba2 (X-SPUR)} & 98.5M & \textbf{0.9987} \\
Transformer             & 97.6M & 0.9944 \\
LSTM                    & 97.6M & 0.9818 \\
Dilated TCN             & 99.4M & 0.9960 \\
\bottomrule
\end{tabular}
\end{table}

\subsection{Computational and Deployment Feasibility}
\label{subsec:deployment_feasibility}
We examine sequence-length scaling using FP32 forward-pass timings on a single RTX 5090.
The parameter counts of Mamba2 and the Transformer differ by less than 1\%.
We use randomly initialized weights and synthetic inputs.
At batch 32, Mamba2 takes 40.57\,ms at $L=256$ and 167.75\,ms at $L=1024$.
The Transformer takes 37.98\,ms and 214.55\,ms, respectively.
At batch 32, Mamba2 is faster at $L=1024$ but not at $L=256$.
The Transformer is faster at both lengths with batch 1.

We also profile the trained 98.5M-parameter model on Jetson AGX Orin 64\,GB in MAXN mode. We use FP16, CUDA Graphs, and synthetic inputs at $L=256$. Timing covers input transfer, model inference, and dual top-$k$ scoring. Mean latency is 6.19\,ms at batch 1 and 140.51\,ms at batch 32. The latter yields 227.74 windows/s with 1.48\,GB of peak PyTorch-reserved GPU memory.
The TOW-IDS captures generate about 1,973 windows/s on average at $w=10$ and stride 1.

To reduce this load, we shorten the input to $L=192$ without retraining.
We retain $w=10$ but evaluate one window every 12 packets within each flow.
At this stride, processing the TOW-IDS traffic requires about 164 window evaluations/s on average.
On Orin, the measured batch-32 throughput is 298.71 windows/s.
This exceeds the average processing requirement.

We also assess detection performance under this sampling policy.
We reduce the smoothing width to $K=7$ for the wider sampling interval.
We recalibrate using benign validation data.
For windows not evaluated by the model, we carry forward the most recent score.
This yields an offline TOW-IDS AUC of 0.9962 over all test windows.
Thus, this setting can meet the average processing demand while retaining high detection performance.

\section{Conclusion}
\label{sec:conclusion}
This article presented X-SPUR, an explainable unsupervised intrusion detection framework for automotive Ethernet that replaces handcrafted multimodal feature engineering with causal next-token prediction over tokenized packet fields. Learning benign sequential structure with a Mamba2 state-space language model, X-SPUR detects intrusions without attack labels while keeping anomaly evidence at the token level.

On the TOW-IDS dataset, X-SPUR achieves an AUC of 0.9987. AERO~\cite{Jeong2024AERO} reports 0.9969. Our ablations show that bimodal payload--timing fusion adds discriminative signal beyond packet content. Flow-level smoothing with dual top-$k$ per-protocol $Z$-score calibration also suppresses protocol-specific false positives to 0.81\% of benign windows. Token-level surprisal yields intrinsic field-level attribution. It exposes distinct attack signatures and traces detections and residual false positives to the same IEC\,61883/MPEG-TS stream fields. We also apply the same model architecture and training hyperparameters to a separately trained CarDS model. It preserves strong detection with an AUC of 0.9923. A faithful AERO reimplementation achieves an AUC of 0.6239. These results support the transferability of the proposed methodology across datasets and protocol stacks.

\BfPara{Limitations and Future Work} AVTP Frame Injection remains the hardest attack to detect.
Its counter and timestamp fields are difficult to predict even in benign traffic.
This leaves a weak surprisal contrast after injection.
The centered $K=63$ smoother uses up to 31 subsequent windows from the same flow.
At stride 1, the median waiting time is 466\,ms on TOW-IDS and 10.7\,ms on CarDS.
The corresponding 95th-percentile values are 959\,ms and 25.2\,ms.
Causal smoothing is a future direction for latency-sensitive deployment.
The 98.5M-parameter model also remains computationally demanding for production ECUs.
The single-Orin profile does not sustain the observed aggregate stride-1 load.
Distillation and quantization remain priorities for embedded deployment.
Field-level surprisal indicates deviation from learned benign behavior.
It supports analyst interpretation but does not establish maliciousness.

\bibliographystyle{IEEEtran}
\bibliography{references}

\begin{IEEEbiography}[{\includegraphics[width=1in,height=1.25in,clip,keepaspectratio]{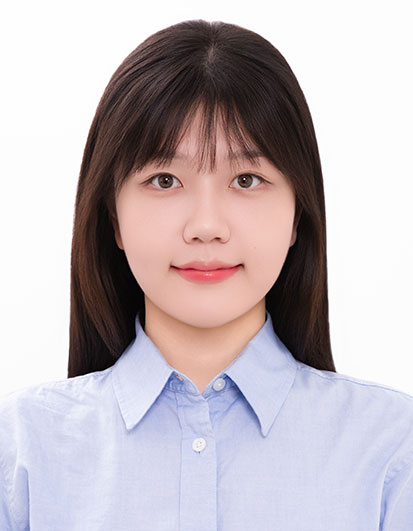}}]{Jisoo Kim}
(Student Member, IEEE) is currently pursuing the B.S. degree in data science with Sookmyung Women's University, Seoul, Republic of Korea.

She is an Undergraduate Researcher with the System and Network Security Laboratory (SNSec Lab), Sookmyung Women's University. Her research focuses on data-driven cybersecurity, particularly explainable and unsupervised intrusion detection for automotive Ethernet. Her research interests include network security, automotive cybersecurity, anomaly detection, and explainable machine learning for cybersecurity.
\end{IEEEbiography}

\begin{IEEEbiography}[{\includegraphics[width=1in,height=1.25in,clip,keepaspectratio]{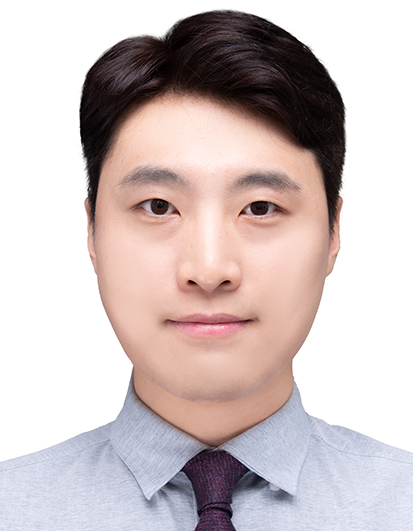}}]{Seonghoon Jeong}
(Member, IEEE) received the Ph.D. degree in information security from Korea University, Seoul, Republic of Korea, in 2023.

He is currently an Assistant Professor with the Division of Artificial Intelligence Engineering, Sookmyung Women's University, Seoul, Republic of Korea. He leads the System and Network Security Laboratory (SNSec Lab). His research focuses on system and network security, particularly explainable and unsupervised intrusion detection for connected vehicles. His current research interests also include learning-based binary analysis, microarchitectural security, and foundation models for cybersecurity.
\end{IEEEbiography}

\end{document}